\documentclass[journal]{IEEEtran}
\usepackage{hyperref}

\usepackage{cite}

\ifCLASSINFOpdf
  \usepackage[pdftex]{graphicx}
  \graphicspath{{./figures/}}
\else
\fi
\usepackage{amsmath}
\usepackage{dsfont}
\usepackage{textcomp}

\usepackage{algorithm}
\usepackage{algpseudocode}

\usepackage{array}

\usepackage{xcolor}

\usepackage{url}

\begin{document}
\title{Physics-based Learned Design: Optimized Coded-Illumination for Quantitative Phase Imaging}
%
%
%

\author{Michael~R.~Kellman,
        Emrah~Bostan,
        Nicole~Repina,
        and~Laura~Waller
\thanks{M. Kellman, E. Bostan, L. Waller are with the Department
of Electrical Engineering and Computer Sciences, University of California, Berkeley,
CA, 94720, USA.}
\thanks{N. Repina is with the Graduate Program in Bioengineering, University of California, Berkeley, CA, 94720, USA.}
}

%
%

\markboth{Journal of \LaTeX\ Class Files,~Vol.~14, No.~8, August~2015}%
{Shell \MakeLowercase{\textit{et al.}}: Bare Demo of IEEEtran.cls for IEEE Journals}
%



\maketitle

\begin{abstract}
Coded-illumination can enable quantitative phase microscopy of transparent samples with minimal hardware requirements. Intensity images are captured with different source patterns and a non-linear phase retrieval optimization reconstructs the image. The non-linear nature of the processing makes optimizing the illumination pattern designs complicated. Traditional techniques for experimental design (\textit{e.g.} condition number optimization, spectral analysis) consider only linear measurement formation models and linear reconstructions. Deep neural networks (DNNs) can efficiently represent the non-linear process and can be optimized over via training in an end-to-end framework. However, DNNs typically require a large amount of training examples and parameters to properly learn the phase retrieval process, without making use of the known physical models. Here, we aim to use both our knowledge of the physics and the power of machine learning together. We develop a new data-driven approach to optimizing coded-illumination patterns for a LED array microscope for a given phase reconstruction algorithm. Our method incorporates both the physics of the measurement scheme and the non-linearity of the reconstruction algorithm into the design problem. This enables efficient parameterization, which allows us to use only a small number of training examples to learn designs that generalize well in the experimental setting without retraining. We show experimental results for both a well-characterized phase target and mouse fibroblast cells using coded-illumination patterns optimized for a sparsity-based phase reconstruction algorithm. Our learned design results using $2$ measurements demonstrate similar accuracy to Fourier Ptychography with $69$ measurements.
\end{abstract}

\begin{IEEEkeywords}
Phase Imaging, Unrolled Network, Physics-based, Experimental Design, Illumination Design.
\end{IEEEkeywords}

%
\IEEEpeerreviewmaketitle

\section{Introduction}
\label{sec:intro}

    \begin{figure*}[t]
        \centering
        \includegraphics[width=13.25cm]{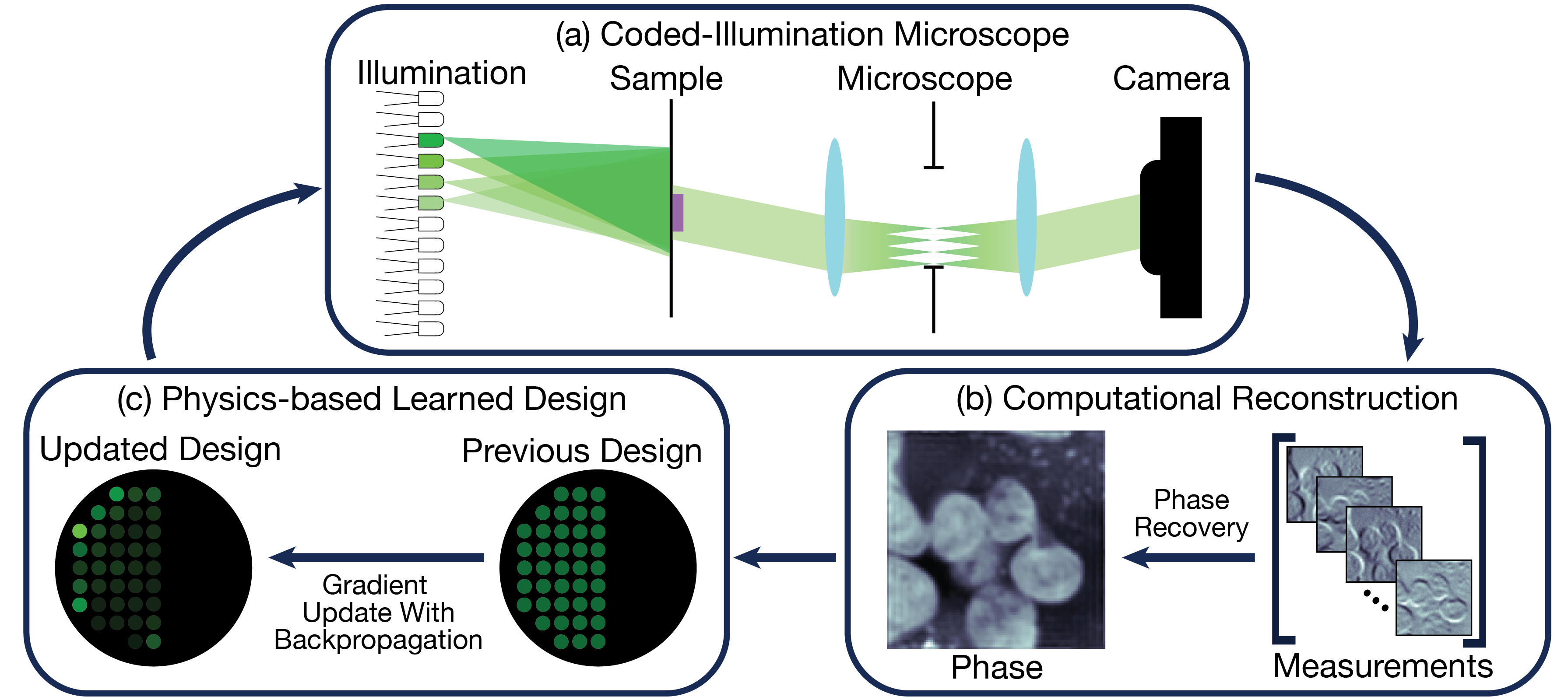}
        \caption{Learning coded-illumination designs for quantitative phase imaging: (a) The LED-array microscope captures multiple intensity measurements with different coded-illumination source patterns. (b) The measurements are used to computationally reconstruct the sample's complex-field using an iterative phase recovery algorithm. (c) An optimization procedure for learning optimal coded-illumination patterns updates the illumination design.}
        \label{fig:fig1}
    \end{figure*}

    Quantitative Phase Imaging (QPI) enables stain-free and label-free microscopy of transparent biological samples \textit{in vitro}~\cite{popescu2011quantitative,Mir:2012jp}. When compared with coherent methods~\cite{Rappaz.etal2014,cuche1999simultaneous}, QPI methods that use partially coherent light achieve higher spatial resolution, more light throughput, and reduced speckle artifacts. Phase contrast may be generated using interference~\cite{Bhaduri:2012vu,Wang.etal2011} or defocus~\cite{Gureyev:95,Streibl:1984we,Waller.etal2010}. More recently, \textit{coded-illumination microscopy}~\cite{Tian:2015ut,Zheng.etal2013,zheng2011microscopy,Tian:2014wv,Tian.Waller2015,Ling:2018vf} has been demonstrated as an accurate and inexpensive QPI scheme. To realize coded-illumination, we replace a commercial microscope's illumination unit with a light-emitting diode (LED) domed array (see Fig.~\ref{fig:fig1})~\cite{Phillips:17}. This provides a flexible hardware platform for various QPI applications including super-resolution~\cite{Tian:2015ut,Zheng.etal2013,Tian:2014wv}, multi-contrast~\cite{zheng2011microscopy,liu2014real}, and 3D imaging~\cite{Tian.Waller2015,Ling:2018vf}.
    

    Coded-illumination microscopy uses asymmetric source patterns~\cite{Kachar766} and multiple measurements to retrieve 2D phase information. Quantitative Differential Phase Contrast~\cite{Hamilton_Sheppard_1984,Mehta:09,Tian:2015fs,Claus:15} (qDPC), for example, captures four measurements with rotated half-circle source patterns, from which the phase is computationally recovered using a partially coherent linearized model. The practical performance of qDPC is predominantly determined by how the phase information is encoded in (via coded-illumination patterns) and decoded from (via phase recovery) the intensity measurements. 
    
    The half-circle illumination designs of qDPC were derived analytically based on a Weak Object Approximation~\cite{Hamilton:1984,Streibl:85,Mehta:09,Tian:2015fs} which linearizes the physics in order to make the inverse problem mathematically convenient. This linearized model enables one to derive a phase transfer function and analyze the spatial frequency coverage of any given source pattern~\cite{Tian:2015fs,Claus:15,Li:17,Lin:2018ks}; however, the non-linearity of the exact model makes it impossible to predict an optimal source design without knowing the sample's phase \textit{a priori}. In addition, these types of analysis are inherently restricted to linear reconstruction algorithms and will not necessarily result in improved accuracy when the phase is retrieved via non-linear iterative methods. 
    
   Motivated by the success of deep learning \cite{LeCun:2015dt} for image reconstruction problems \cite{Jin:2016up,Wangetal.2016,Rivenson:ksa,Sinha:17,Rivenson:17,DBLP:journals/corr/abs-1805-00334}, data-driven approaches have been adopted for learning coded-illumination patterns. For instance, researchers have used machine learning to maximize the phase contrast of each coded-illumination measurement \cite{Diederich:2018ht}, to improve accuracy on classification tasks \cite{Horstmeyer:2017tg}, and to reconstruct phase \cite{Robeyetal.2018}. All of these techniques learn the input-output relationship with a deep convolutional neural network (CNN) using training data. It is not straightforward to include the well-characterized system physics; hence, the CNN is required to learn both the physical measurement formation and the phase reconstruction process. This task requires training of 10s to 100s of thousands of parameters and an immense number of training examples.

    Here, we introduce a new data-driven approach to optimizing the source pattern design for coded-illumination phase retrieval by directly including both the system physics and the non-linear nature of a reconstruction algorithm in the learning process. Our approach \textit{unrolls} the iterations of a generic non-linear reconstruction algorithm to construct an \textit{unrolled network}~\cite{gregor2010learning,Hammernik:2017ku,Diamond:2017wa,sun2016deep,Anonymous:XgcAjXu7,Bostan:2018cr}. Similar to CNNs, our \textit{unrolled network} consists of several layers (one for each iteration); however, in our case each layer consists of well-specified operations to incorporate measurement formation and sparse regularization, instead of standard operations such as generic convolutions. The key aspects of our approach are:

    \begin{itemize}
        \item incorporation of the system physics and reconstruction non-linearities in the illumination design process.
        \item efficient parameterization of the unrolled network.
        \item incorporation of practical constraints.
        \item reduced number of training examples required.
    \end{itemize}

We deploy our data-driven approach to learn improved coded-illumination patterns for phase reconstruction. Each layer of the unrolled network is parameterized by only a few variables (LED brightness values), enabling an efficient use of training data ($<100$ simulated training examples). We compare the QPI performance of our learned designs to previous work and demonstrate that our designs generalize well to the experimental setting with biological samples.

\section{Quantitative Phase Imaging}
\label{sec:qpi}

qDPC recovers a sample's complex transmittance function from several coded-illumination measurements. The phase recovery optimization algorithm aims to minimize the Euclidean norm of the error between the measurements and the expected measurements formed with the current phase estimate. Using a gradient-based procedure, the phase estimate is iteratively updated until convergence. For a partially coherent source, the phase can be recovered with resolution up to twice the coherent diffraction limit. In this section, we describe the measurement formation process and phase recovery optimization.

\subsection{System Modelling}
    \label{ssec:model}

A thin sample's transmission function can be approximated as a 2D complex function, $o(\mathbf{r})=e^{j\phi(\mathbf{r}) - \mu(\mathbf{r})}$, characterized by its absorption, $\mu(\mathbf{r})$, and phase, $\phi(\mathbf{r}) = \frac{2\pi}{\lambda}\Delta n(\mathbf{r}) d(\mathbf{r})$, where $\mathbf{r}$ are 2D spatial coordinates, $\lambda$ is the wavelength of the illumination, $d(\mathbf{r})$ is the physical thickness of the sample, and $\Delta n(\mathbf{r})$ is the change in refractive index from the background. Intensity measurements, $y(\mathbf{r})$, of the sample are a non-linear function of $o(\mathbf{r})$, mathematically described by,

    \begin{align}
        y(\mathbf{r}) = |p(\mathbf{r})*(s(\mathbf{r}) \odot o(\mathbf{r}))|^2,
        \label{eq:eq01}
    \end{align}
        
\noindent where $|\cdot|^2$ denotes squared absolute value, $*$ denotes convolution, $\odot$ denotes elementwise multiplication, $s(\mathbf{r})$ is the illumination's complex-field at the sample plane and $p(\mathbf{r})$ is the point spread function (PSF) of the microscope. The illumination from each LED is approximated as a tilted plane wave, $s(\mathbf{r}) = e^{\frac{j}{\lambda}\mathbf{u}_{pos}^T\mathbf{r}}$, with tilt angle, $\mathbf{u}_{pos}$, determined by the physical position of the LED relative the microscope~\cite{Zheng:2013gq}.

Because the measured image in Eq.~\ref{eq:eq01} is non-linear with respect to the sample's transmission function, recovering phase generally requires non-convex optimization. However, biological samples in closely index-matched fluid have a small \textit{scatter-scatter} term. This means that a \textit{weak object approximation} can be made; linearizing the measurement formation model such that phase recovery requires only a linear deconvolution of the measurements with their respective weak object transfer functions (WOTFs)~\cite{Hamilton:1984,Mehta:09,Streibl:85,Claus:15,Tian:2015fs}. Further, unstained biological samples are predominantly phase objects since they are only weakly absorbing (\textit{i.e.} $\mu(\mathbf{r})$ is small). With these approximations, we can express each intensity measurement as a linear system with contributions from the background and phase contrast. In Fourier space, 
        
    \begin{align}\label{eq:weakObjectApprox}
        \widehat{y}(\mathbf{u}) \approx B\delta(\mathbf{u}) + i h(\mathbf{u})\widehat\phi(\mathbf{u}),
    \end{align}

\noindent where $\widehat{\cdot}$ denotes Fourier transform, $\mathbf{u}$ are 2D spatial-frequency coordinates, $B$ is the measurement's background energy concentrated at the DC and $h(\mathbf{u})$ is the phase WOTF. The phase WOTFs are a function of the illumination source and the pupil distribution of the microscope~\cite{Tian:2015fs}. For a single LED the WOTF is:

        \begin{align}
            h^{(single)}(\mathbf{u}) &= i(\widehat{p}(\mathbf{u}) \star \widehat{s}(\mathbf{u}) - \widehat{s}(\mathbf{u}) \star \widehat{p}(\mathbf{u})),
        \end{align}

\noindent where $\star$ is the correlation operator, defined as $(x_1 \star x_2)(\mathbf{r}) = \int x_1(\tilde{\mathbf{r}}) x_2^{*}(\tilde{\mathbf{r}}-\mathbf{r}) d\tilde{\mathbf{r}}$ for $\mathbf{r}$ in the domain of $\widehat{p}$ and $\widehat{s}$.
        
In~\cite{Tian:2015fs}, multiple LEDs are turned on simultaneously to increase signal-to-noise (SNR) and improve phase contrast. Because the fields generated by each LED's illumination are spatially incoherent with each other, the measurement from multiple LEDs will simply be the weighted sum of each LED's individual measurement, where the weights correspond to the LEDs' brightness values. The phase WOTF for illumination by multiple LEDs will also be the weighted sum of the single-LED phase WOTFs. Mathematically, 
\begin{align}
    \widehat{y}^{(multi)}(\mathbf{u}) &= \sum_{w \in \mathcal{W}} c_w \widehat{y}^{(single)}(\mathbf{u}) \\
            h^{(multi)}(\mathbf{u}) &= \sum_{w \in \mathcal{W}} c_w h_w^{(single)}(\mathbf{u}),
        \end{align}

\noindent where $\mathcal{W}$ is the set of LEDs turned on and $c_w \geq 0$ are the LEDs' brightness values. 

Following common practice~\cite{Bostan.etal2013}, we discretize the 2D spatial distributions and format them as vectors (bold lower case) (\textit{e.g.} $\widehat{\mathbf{h}}$ represents the transfer function's 2D spatial-frequency distribution and $\boldsymbol\phi$ represents the 2D spatial phase distribution). The measurements\footnote{In practice, $\mathbf{y}$ typically refers to the so-called flattened image, where the background energy in~\eqref{eq:weakObjectApprox} is removed via background subtraction.} are described in Fourier space as $\widehat{\mathbf{y}} = \mathbf{A}\widehat{\boldsymbol{\phi}}$ with system function $\mathbf{A} = diag(\widehat{\mathbf{h}})$.

Based on this model, we define $\mathbf{Y} \in \mathds{R}^{M \times S}$ as the Fourier transform of $S$ single LED measurements, $\widehat{\mathbf{y}}$, along the columns. Then, $\mathbf{C} \in \mathds{R}^{S \times K}$ is defined as the $S$ single-LED weights for each of $K$ measurements, and $\mathbf{c}_k \in \mathds{R}^{S}$ is the $k^{th}$ column of $\mathbf{C}$. The product $\widehat{\mathbf{y}}_k = \mathbf{Y}\mathbf{c}_k$ simulates the $k^{th}$ multiple-LED measurement. Similarly, we define $\mathbf{H} \in \mathds{R}^{N \times S}$ as $S$ single LED phase WOTFs, $\widehat{\mathbf{h}}$ along the columns, such that the product $\mathbf{A}_k = diag(\mathbf{H}\mathbf{c}_k)$ gives the corresponding multiple-LED phase WOTF for the $k^{th}$ measurement.

\subsection{Phase Recovery}
\label{ssec:inverseProblem}

Phase recovery using the forward model in Sec.~\ref{ssec:model} can be formulated as a regularized linear inverse problem,

    \begin{align}
        \widehat{\boldsymbol\phi}^{\star} &= \mathcal{R}((\widehat{\mathbf{y}}_k)_{k=1}^{K}, \mathcal{P}(\cdot)) \\
        &= \arg \underset{\widehat{\boldsymbol\phi}}{ \min} \,\, \frac{1}{2K}\sum_{k=1}^{K} \|\widehat{\mathbf{y}}_k - \mathbf{A}_k\widehat{\boldsymbol{\phi}}\|_2^2 + \mathcal{P}(\widehat{\boldsymbol{\phi}}), \label{eq:invproblem}
    \end{align}
        
\noindent where $\boldsymbol{\phi}^{\star}$ is the recovered phase, $K$ is the number of measurements acquired, $\mathbf{\widehat{y}}_k$ is the Fourier transform of the $k^{th}$ measurement and $\mathcal{P}(\cdot)$ is a user-chosen regularizer. 
We solve this optimization problem efficiently using the accelerated proximal gradient descent (APGD) algorithm by iteratively applying an acceleration update, a gradient update and a proximal update~\cite{Parikh:2013vb,Beck:2009gh}. The algorithm is detailed in Alg.~\ref{alg:APGD}, where $\alpha$ is the gradient step size, $N$ is the number of iterations, $\mathbf{s}$ and $\mathbf{z}$ are intermediate variables, $\mu^{(n)}$ is the acceleration parameter derived by the recursion, $\mu^{(n)}=\frac{1 + \sqrt{1 + 4\mu^{(n-1),2}}}{2}$~\cite{Beck:2009gh}, and $\text{prox}_{\mathcal{P}}(\cdot)$ is the proximal operator corresponding to the user-chosen regularizer $\mathcal{P}(\cdot)$~\cite{Parikh:2013vb}.

\begin{figure*}[tbh]
    \centering
    \includegraphics[width=12cm]{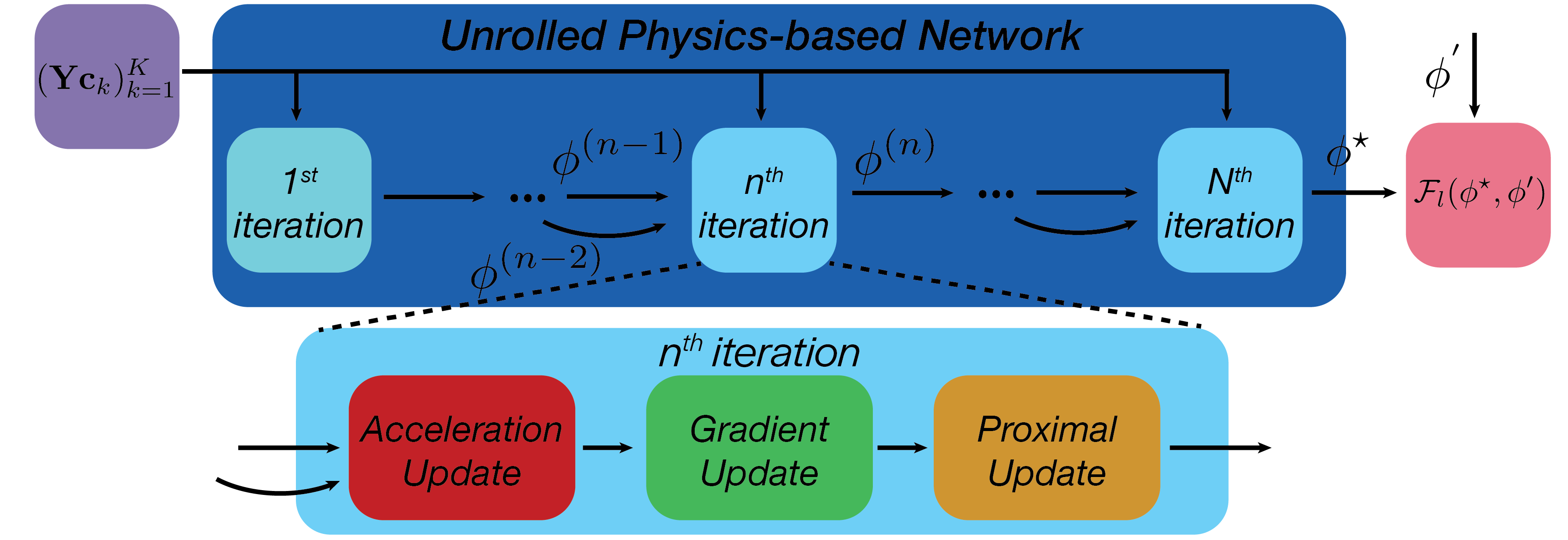}
    \caption{Unrolled physics-based network: Feed-forward schematic for the unrolled accelerated proximal gradient descent (APGD) network for $N$ iterations (dark blue box). The network takes intensity measurements, $\mathbf{y}_k$, parameterized by the coded-illumination design, $\mathbf{c}_k$, as input and outputs the reconstructed phase, $\phi^{\star}$. Finally, the output is compared with the ground truth phase, $\phi'$, using a user-chosen loss function, $\mathcal{F}_l$ (pink box). The inset into a single ($n^{th}$) iteration (light blue box) shows each iteration's three steps: acceleration update, gradient update, and proximal update.}
    \label{fig:fig3}
\end{figure*}
    
\begin{algorithm}[tbh]
\caption{Accelerated Proximal Gradient Descent (APGD) for Phase Recovery} \label{alg:APGD}
\begin{algorithmic}[1]

\Procedure{APGD}{$(\mathbf{\widehat{y}}_k)^{K}_{k=1},N,\alpha, \mathcal{P}(\cdot)$}
    \State $\widehat{\boldsymbol{\phi}}^{(0)} = \mathbf{0}, \widehat{\boldsymbol{\phi}}^{(-1)} = \mathbf{0}$
    \For{$n \in \{1 ... N\}$}
    \State $\mathbf{s}^{(n)} \gets \mu^{(n)}\widehat{\boldsymbol{\phi}}^{(n-1)} + (1-\mu^{(n)})\widehat{\boldsymbol{\phi}}^{(n-2)}$
            \State $\mathbf{z}^{(n)} \gets \mathbf{s}^{(n)} - \frac{\alpha}{K}\sum_{k=1}^K(-\mathbf{A}_k^{H})(\mathbf{\widehat{y}}_k - \mathbf{A}_k\mathbf{s}^{(n)})$
            \State $\widehat{\boldsymbol{\phi}}^{(n)} \gets \text{prox}_{\alpha \mathcal{P}}(\mathbf{z}^{(n)})$
            \EndFor
    \State \textbf{return} $\widehat{\boldsymbol{\phi}}^{(N)}$ \EndProcedure
    \end{algorithmic}
\end{algorithm}

\section{Physics-Based Learned Design}
\label{sec:learning}

Given the phase recovery algorithm in Sec.~\ref{ssec:inverseProblem}, we now describe our main contribution of learning the coded-illumination designs for a given reconstruction algorithm and training set.

\subsection{Unrolled Physics-based Network}
\label{ssec:unrolledNetwork}

Traditionally, DNNs contain many layers of weighted linear mixtures and non-linear activation functions~\cite{LeCun:2015dt}. Here, we consider specific linear functions which capture the system physics of measurement formation and specific non-linear activation functions which promote sparsity. Starting from Alg.~\ref{alg:APGD}, we treat each iteration as a layer such that when unrolled they form a network of $N$ layers, denoted $\mathcal{R}$ (Fig.~\ref{fig:fig3}). Each layer of $\mathcal{R}$ contains a module for each of the iterative algorithm's updates (\textit{i.e.} an acceleration module, a gradient module (incorporates system physics), and a proximal module (incorporates sparsity)). The regularization and step size parameters specified for Alg.~\ref{alg:APGD} are fixed. The network's inputs comprise $(\widehat{\mathbf{y}}_k)_{k=1}^{K}$ and the network's output is $\widehat{\boldsymbol{\phi}}^{(N)}$. The design parameters of the network, which will be learned, govern the relative brightness of the LEDs and are incorporated in the measurement formation and the system WOTFs.

\subsection{Learning Objective}
\label{ssec:learn_objective}
        
Our learning objective is to minimize the phase reconstruction error of the training data over the space of possible LED configurations, subject to constraints that enforce physical feasibility and eliminate degenerate and trivial solutions: 
    
        \begin{align}
            \mathbf{C}^{\star} = & \arg \underset{\mathbf{C}}{ \min} \ \mathcal{F}(\mathbf{C}) \label{eq:cost} \\
            \text{s.t.} \ \  & \mathbf{c}_k \geq 0 \,\,   & \text{(non-negativity)} \label{eq:positive} \\
            & \|\mathbf{c}_k\|_1 = 1 \,\,  & \text{(scale)}\label{eq:power}  \\
            & \mathbf{m}_k \odot \mathbf{c}_k = \mathbf{0}  \,\,  & \text{(geometric)} \label{eq:phasecon} \\
            & \forall k \in \{1 \hdots K\} \nonumber, 
        \end{align}
        \noindent where,

        \begin{align}
            \mathcal{F}(C) &= \frac{1}{L} \sum_{l=1}^{L} \mathcal{F}_l(\mathbf{C}) \\
            &= \frac{1}{2L} \sum_{l=1}^{L} \|\mathcal{R}((\mathbf{Y}_l\mathbf{c}_k)_{k=1}^{K}) - \widehat{\boldsymbol{\phi}}'_l\|_2^2.
        \end{align}

\noindent Here, $(\mathbf{Y}_l, \boldsymbol{\phi}'_l)_{l=1}^{L}$ are $L$ training pairs for which $\mathbf{Y}_l$ is a matrix of the Fourier transform of single-LED measurements for the $l^{th}$ sample with optical phase, $\boldsymbol{\phi}'_l$. $\odot$ is the elementwise product operator, $\mathbf{m}_k$ is a geometric constraint mask for the $k^{th}$ measurement, and $\mathbf{0}$ is the null vector.

The non-negativity constraint (Eq.~\ref{eq:positive}) prevents non-physical solutions by enforcing the brightness of each LED to be greater than or equal to zero. This is enforced by projecting the parameters onto the set of non-negative real numbers. The scale constraint (Eq.~\ref{eq:power}) enforces that each coded-illumination design must have weights with sum equal to 1, in order to eliminate arbitrary scalings of the same design. This is enforced by scaling the parameters for each measurement such that their sum is one. The geometric constraint (Eq.~\ref{eq:phasecon}) enforces that the coded-illumination designs do not use conjugate-symmetric LED pairs to illuminate the sample within the same measurement, since these would also result in degenerate solutions (\textit{e.g.} two symmetric LEDs produce opposite phase contrast measurements that would cancel each other out). To prevent this, we force the source patterns for each measurement to reside within only one of the major semi-circle sets (\textit{e.g.} top, bottom, left, right). This constraint is enforced by setting the LED brightnesses outside the allowed semi-circle to zero.

We solve Eq.~\ref{eq:cost} iteratively via accelerated projected gradient descent (Alg.~\ref{alg:CLA}). At each iteration, the coded-illumination design for each measurement is updated with the analytical gradient, projected onto the constraints (denoted by $\mathcal{B}(\cdot)$) and updated again with a contribution from the previous iteration (weighted by $\beta^{(t)}$). $\mathcal{B}(\cdot)$ enforces the constraints in the following order: non-negativity, geometric, and scale.
    
        \begin{algorithm}[H]
            \caption{Physics-based Learned Design Algorithm }\label{alg:CLA}
            \begin{algorithmic}[1]
            \Procedure{PBLD}{$(\mathbf{Y}_l,\boldsymbol{\phi}'_l)^{L}_{l=0},\mathbf{C},\gamma, T$}
            \For{$t \in \{0 ... T\}$} \Comment{Gradient descent loop}
            \For{$l \in \{1 ... L\}$} \Comment{Training data loop}
            \State $r_l \gets \mathcal{R}((\mathbf{Y}_l\mathbf{c}_k)_{k=1}^{K}) - \widehat{\boldsymbol{\phi}}'_l$
            \State $\mathbf{G}_l \gets \textit{BackPropagation}(r_l)$
            \EndFor
            \State $\mathbf{C}^{(t+1)} \gets \mathcal{B}(\mathbf{C}^{(t)}  - \frac{\gamma}{L} \sum_{l=1}^{L} \mathbf{G}_l)$ 
            \State $\mathbf{C}^{(t+1)} \gets \beta^{(t)}\mathbf{C}^{(t+1)} + (1-\beta^{(t)})\mathbf{C}^{(t)}$ 
            \EndFor
            \State \textbf{return} $\mathbf{C}^{(T)}$ \EndProcedure
            \end{algorithmic}
        \end{algorithm}

\subsection{Gradient Update}
\label{ssec:gradient}
        
The gradient of the loss function (Eq.~\ref{eq:cost}) with respect to the design parameters has contributions at every layer of the unrolled network through both the measurement terms, $\widehat{\mathbf{y}}_k$, and the phase WOTF terms, $\mathbf{A}_k$, for each measurement $k \in \{1...K\}$. Here, we outline our algorithm for updating the coded-illumination design weights via a two-step procedure: backpropagating the error from layer-to-layer and computing each layer's gradient contribution. For simplicity, we outline the gradient update for only a single training example, $l$, as the gradient for all the training examples is the sum of their individual gradients.

Unlike pure gradient descent, where each iteration's estimate only depends on the previous', accelerated methods like Alg.~\ref{alg:APGD} linearly combine the previous two iteration's estimates to improve convergence. As a consequence, backpropagating error from layer-to-layer requires contributions from two successive layers. Specifically, we compute the error at all $N$ layers with the recursive relation,

        \begin{align}
            \frac{\partial \mathcal{F}_l}{\partial \widehat{\boldsymbol{\phi}}^{(n-2)}} &=  \frac{\partial \mathbf{s}^{(n)}}{\partial \widehat{\boldsymbol{\phi}}^{(n-2)}}\frac{\partial \mathbf{z}^{(n)}}{\partial \mathbf{s}^{(n)}}\frac{\partial \widehat{\boldsymbol{\phi}}^{(n)}}{\partial \mathbf{z}^{(n)}}\frac{\partial \mathcal{F}_l}{\partial \widehat{\boldsymbol{\phi}}^{(n)}} \nonumber \\
            &+ \frac{\partial \mathbf{s}^{(n-1)}}{\partial \widehat{\boldsymbol{\phi}}^{(n-2)}}\frac{\partial \mathbf{z}^{(n-1)}}{\partial \mathbf{s}^{(n-1)}}\frac{\partial \widehat{\boldsymbol{\phi}}^{(n-1)}}{\partial \mathbf{z}^{(n-1)}}\frac{\partial \mathcal{F}_l}{\partial \widehat{\boldsymbol{\phi}}^{(n-1)}},
        \end{align}

\noindent where each partial gradient constitutes a single step in Alg.~\ref{alg:APGD} (fully derived in the supplement).

With the backpropagated error at each layer, we compute the gradient of the loss function with respect to $\mathbf{C}$ as,

        \begin{align}
            \nabla_{\mathbf{C}} \mathcal{F}_l(\mathbf{C}) &= \sum_{n=0}^{N} \mathbf{Q}^{(n)},
        \end{align}

        \noindent for which,

        \begin{align}
            \mathbf{Q}^{(n)} = \frac{\alpha}{K}\sum_{k=1}^{K} (\frac{\partial \mathbf{A}^H_{k} \widehat{\mathbf{y}}_k}{\partial \mathbf{C}} - \frac{\partial \mathbf{A}^H_{k}\mathbf{A}_{k}}{\partial \mathbf{C}}\mathbf{s}^{(n-1)})\frac{\partial \widehat{\boldsymbol{\phi}}^{(n)}}{\partial \mathbf{z}^{(n)}}\frac{\partial \mathcal{F}_l}{\partial \widehat{\boldsymbol{\phi}}^{(n)}}.
        \end{align}

\noindent Here, $ \left( \partial \widehat{\boldsymbol{\phi}}^{(n)} \middle/ \partial \mathbf{z}^{(n)} \right)$ backpropagates the error through the proximal operator and other partials with respect to $\mathbf{C}$ relate the backpropagated error at each layer to the changes in $\mathbf{C}$. Derivations of these partial gradients are included in the supplementary material. In Alg.~\ref{alg:BP}, we unite these two steps to form a recursive algorithm which efficiently computes the analytic gradient for a single training example. Alternatively, general purpose auto-differentiation included in learning libraries (\textit{e.g.} PyTorch, TensorFlow) can be used to perform the gradient updates.

        \begin{algorithm}[H]
            \caption{Gradient Update for Single Training Example}\label{alg:BP}
            \begin{algorithmic}[1]
            \Procedure{Backpropagation(BP)}{$\mathbf{r}^{(N)}$}
            \For{$n \in \{N ... 0\}$} 
            \State $\mathbf{b}^{(n)} \gets \frac{\partial \widehat{\boldsymbol{\phi}}}{\partial \mathbf{z}} \mathbf{r}^{(n)}$
            \State $\mathbf{v}^{(n)} \gets (I-\frac{\alpha}{K}\sum_{k=1}^{K}\mathbf{A}^{H}_{k} \mathbf{A}_{k})\mathbf{b}^{(n)}$
            \State $\mathbf{r}^{(n-1)} \gets \mu^{(n)} \mathbf{v}^{(n)} + (1-\mu^{(n+1)}) \mathbf{v}^{(n+1)}$
            \State  $\mathbf{Q}^{(n)} \gets \frac{\alpha}{K}\sum_{k=1}^{K} (\frac{\partial \mathbf{A}^H_{k} \mathbf{\widehat{y}}_k}{\partial \mathbf{C}} - \frac{\partial \mathbf{A}^H_{k}\mathbf{A}_{k}}{\partial \mathbf{C}}s^{(n-1)})\mathbf{b}^{(n)}$
            \EndFor
            \State \textbf{return} $\sum_{n=0}^{N} \mathbf{Q}^{(n)}$ \EndProcedure
            \end{algorithmic}
        \end{algorithm}

\begin{figure}[h!]
    \centering
    \includegraphics[width=8.89cm]{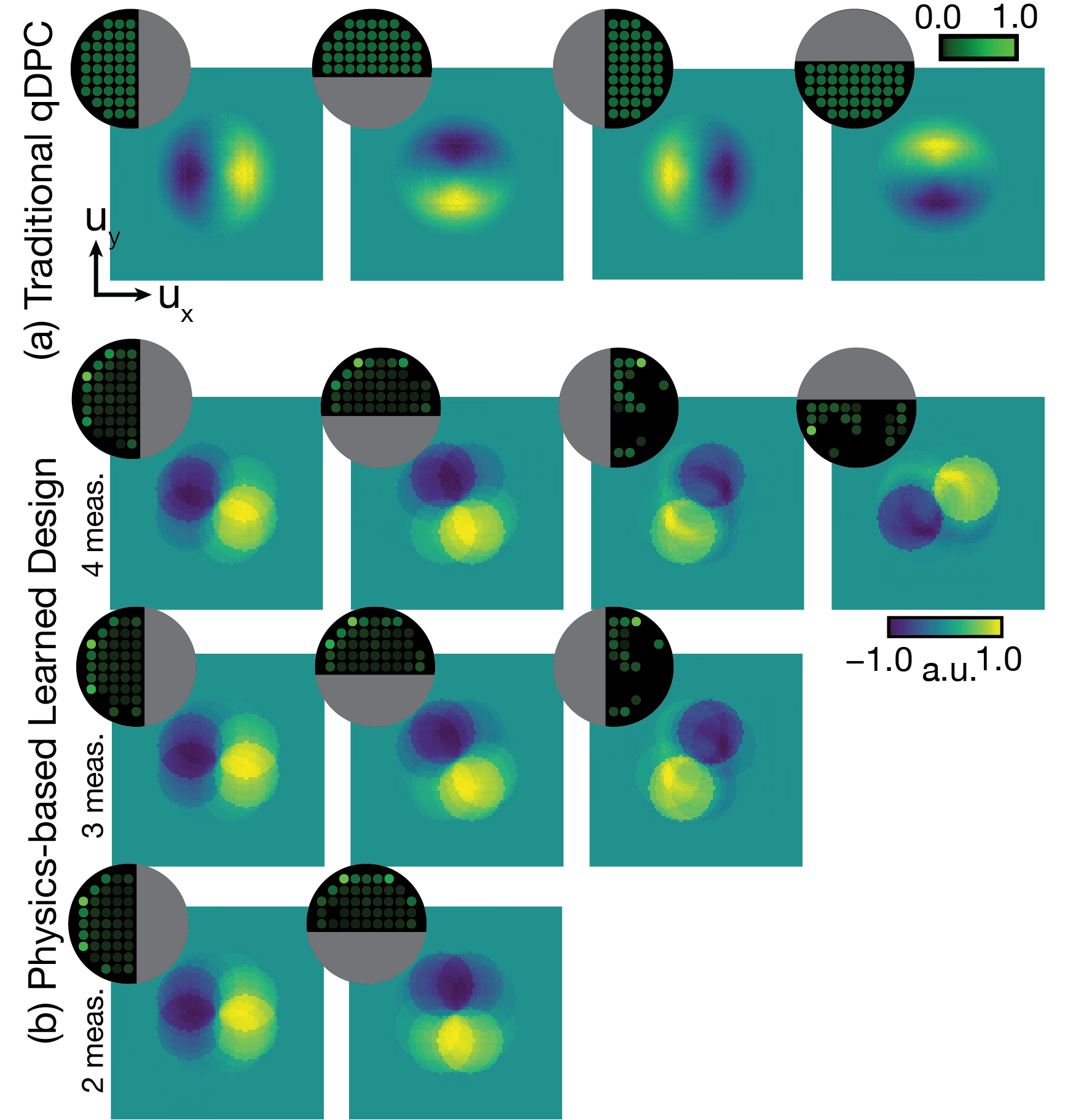}
    \caption{Coded-illumination designs and their corresponding phase weak object transfer functions (WOTFs) for: (a) Traditional qDPC and (b) learned designs for the case where 4, 3, or 2 measurements are allowed for each phase reconstruction. The illumination source patterns are in the upper left corners, with gray semi-circles denoting where the LEDs are constrained to be ``off''.}
    \label{fig:fig4}
\end{figure}

\section{Results}
\label{sec:results}

Our proposed method learns the coded-illumination design for a given reconstruction and training set (Fig.~\ref{fig:fig4}b), yet up to this point we have not detailed specific parameters of our phase reconstruction. In our results, we set the parameters of our reconstruction algorithm (Alg.~\ref{alg:APGD}) to have a fixed CPU time by fixing the number of iterations at $N = 40$ and the step size to $\alpha = 0.2$ (see supplement for parameter analysis). In addition, the regularization term, $\mathcal{P}(\boldsymbol{\phi})$, has been defined generally (\textit{e.g.} $\ell_1$ penalty, total variation (TV) penalty~\cite{osher2005iterative}, BM3D~\cite{dabov2007image}). Here, we choose to enforce TV-based sparsity:

    \begin{align}
        \mathcal{P}(\boldsymbol{\phi}) &= \tau \sum_{i}\|D_i \boldsymbol{\phi}\|_1,
        \label{eq:reg}
    \end{align}

\noindent where $\tau = 1\text{e}^{-3}$ is set to trade off the TV cost with the data consistency cost and $D_i$ is the first-order difference operator along the $i^{th}$ image dimension. We efficiently implement the proximal operator of Eq.~\ref{eq:reg} in closed form via parallel proximal method~\cite{Bostan:2018cr,combettes2011proximal,Kamilov:2016gc} (details in supplement).
    
\subsection{Learning}

\begin{figure*}[t]
    \centering
    \includegraphics[width=18cm]{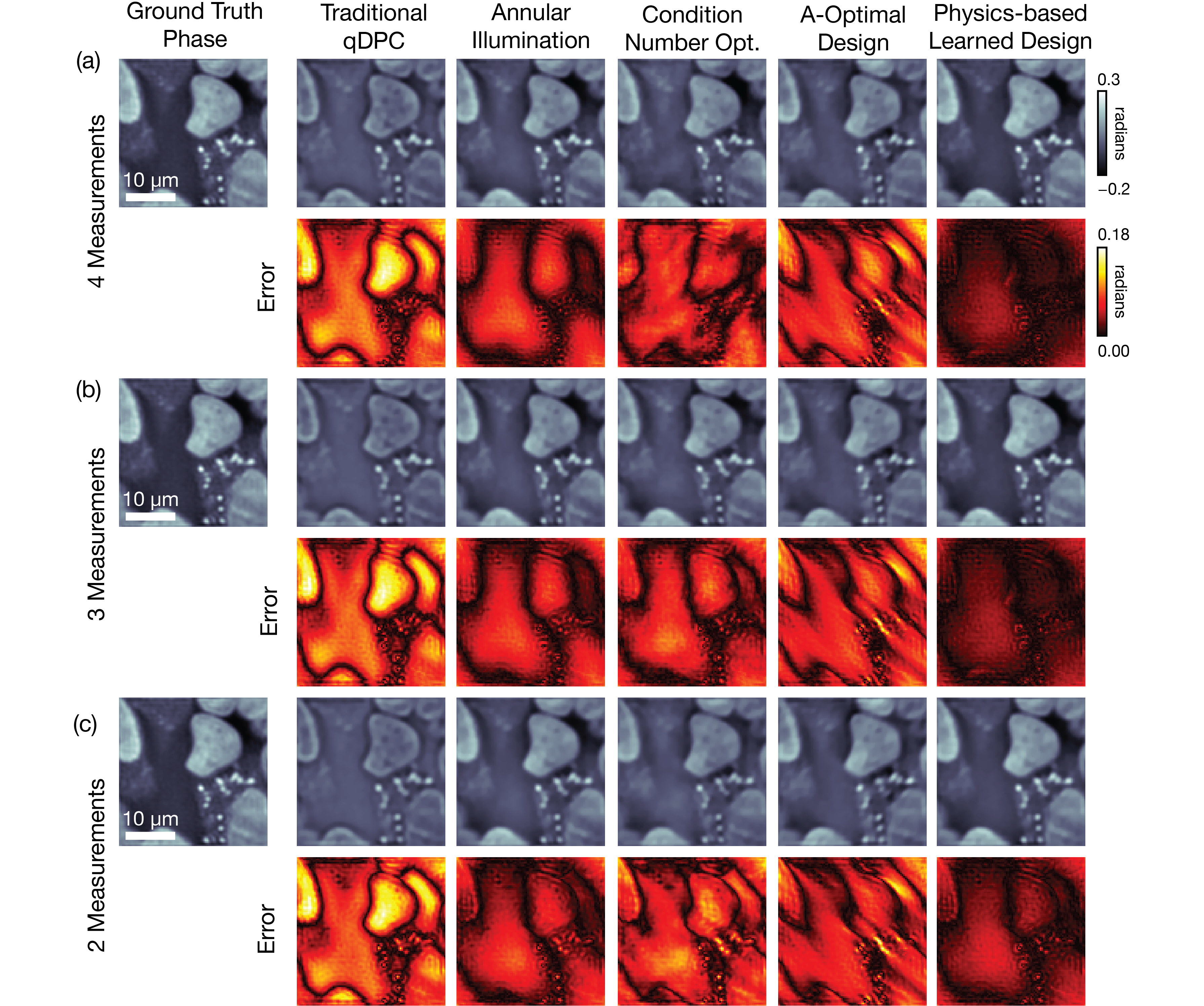}
    \caption{Phase reconstruction results using simulated measurements with different coded-illumination designs. We compare results from: traditional qDPC (half-circles), annular illumination, condition number optimization, A-optimal design, and our proposed physics-based learned designs. We show results for the cases of (a) four, (b) three, and (c) two measurements allowed for each phase reconstruction. Absolute error maps are shown below each reconstruction.}
    \label{fig:fig5}
\end{figure*}

To train our coded-illumination design parameters using Alg.~\ref{alg:CLA}, we generate a dataset of 100 examples (90 for training, 10 for testing). Each example contains ground truth phase from a small region ($95 \times 95\text{pixels}$) of a larger image and $69$ simulated single LED measurements (using Eq.~\ref{eq:eq01}). The LEDs are uniformly spaced within a circle such that each single-LED intensity measurement is a brightfield measurement. The physical system parameters used to generate the phase WOTFs and simulate the training data measurements are $\lambda = 0.532\mu m$, $\text{pixel pitch} = 6.5\mu m$, $\text{magnification} = 20\times$, and $NA_{obj} = 0.25$. To train, we use $\ell_2$ cost between reconstructed phase and ground truth phase as our loss function and approximate the full gradient of Eq.~\ref{eq:cost} with a batch gradient from random batches of $10\%$ of the training pairs at each iteration. We use a learning rate of $\gamma = 1\text{e}^{-2}$ (training and testing convergence curves are provided in the supplement). The training is performed on a multi-core CPU (Dual-socket Intel Xeon\textregistered$\ $ E5 Processor @ 2.1GHz with $64$ cores and $504$GB of RAM) and batch updates are computed in parallel with each training example on a single core. Each batch update takes $\sim$~6 seconds. 200 updates are performed, resulting in a total training time of 20 minutes.

        \begin{table*}[t!]
            \centering
            \caption{PSNR Results: Average and standard deviation PSNR (dB) of phase reconstructions from the simulated testing examples using different illumination schemes and different numbers of measurements. Factor format: Mean $\pm$ Std.}
            \begin{tabular}{ |c||c|c|c|c|c|c| }
                \hline
                \# Meas. & Random & Traditional & Annular & Cond. Number & A-optimal & Physics-based \\
                & Illumination & qDPC & Illumination & Optimization & Design & Learned Design \\
                \hline
                4 & 12.30 $\pm$ 2.12 & 15.67 $\pm$ 2.19 & 20.40 $\pm$ 2.09 & 20.37 $\pm$ 2.41 & 17.94 $\pm$ 2.54 & \bf{28.46} $\pm$ 2.50 \\
                3 & 12.33 $\pm$ 2.12 & 15.28 $\pm$ 2.18 & 20.44 $\pm$ 2.26 & 19.33 $\pm$ 2.03 & 18.05 $\pm$ 2.59 & \bf{28.04} $\pm$ 2.59 \\
                2 & 12.25 $\pm$ 2.12 & 14.87 $\pm$ 2.23 & 20.21 $\pm$ 2.24 & 17.19 $\pm$ 2.28 & 18.08 $\pm$ 2.64 & \bf{23.73} $\pm$ 2.18 \\
                \hline
            \end{tabular}
            \label{table:dBimprovement}
        \end{table*}

        
        
        
\begin{figure*}[tbh]
    \centering
    \includegraphics[width=18.19cm]{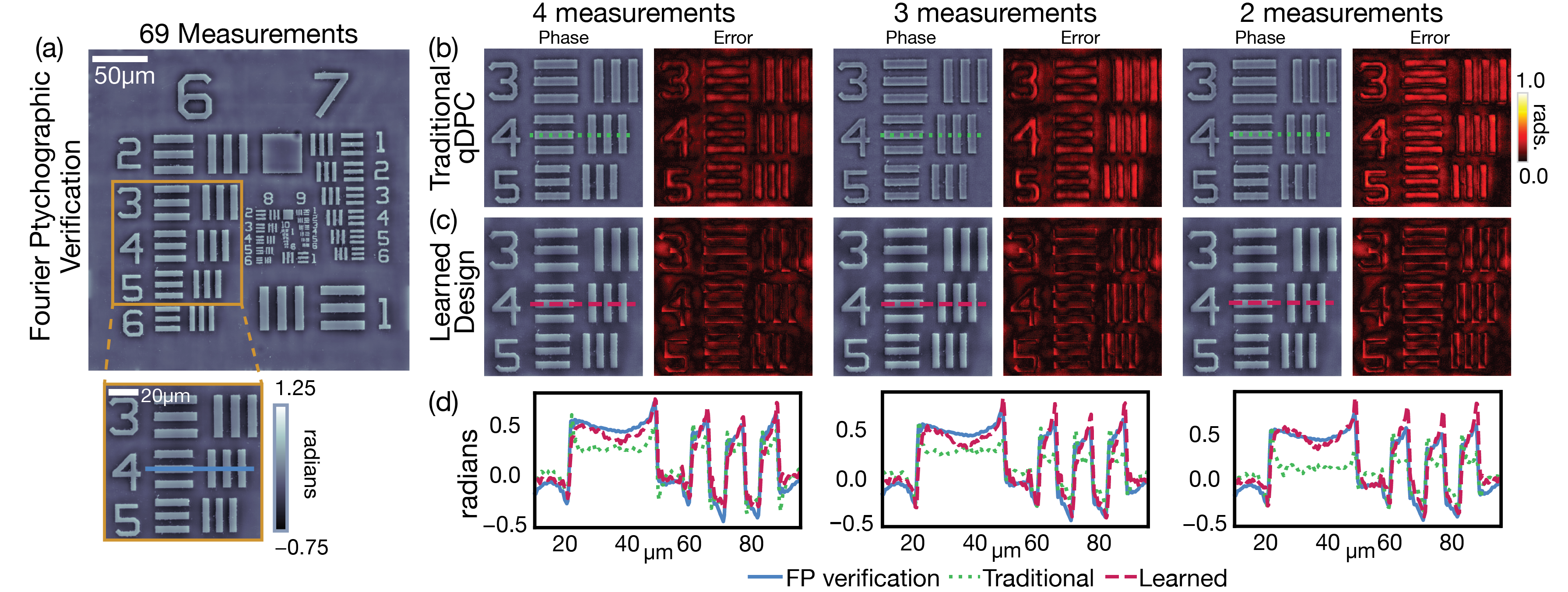}
    \caption{USAF phase target reconstructions: Experimental comparison between phase results with (a) Fourier Ptychography (FP) using 69 images, (b) traditional qDPC and (c) learned designs, for the case of 4, 3, and 2 measurements. Error maps show the difference from the FP reconstruction. (d) Cross-sections show that phase from our learned designs (long-dashed red) is closer to that of FP (solid blue) than traditional qDPC (short-dashed green).}
    \label{fig:fig6}
\end{figure*}
\begin{figure*}[t!]
    \centering
    \includegraphics[width=18.19cm]{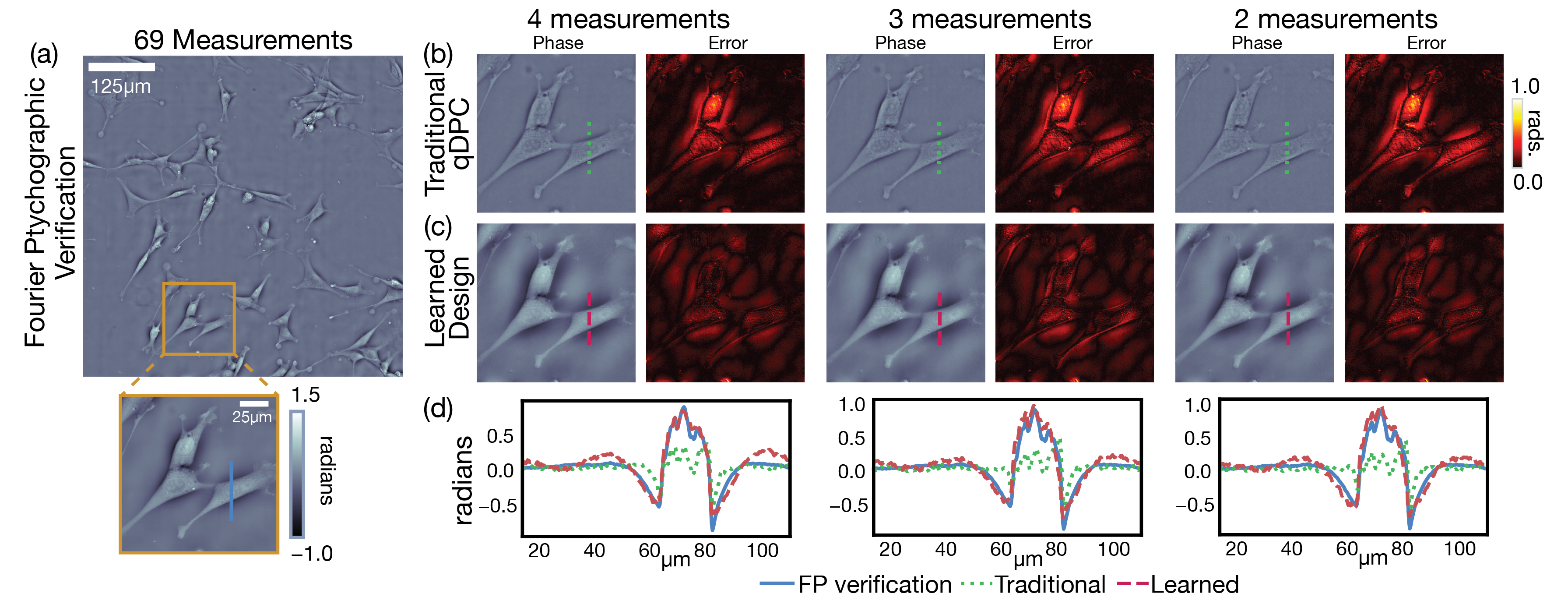}
    \caption{3T3 mouse fibroblast cells reconstructions: Experimental comparison between phase results with (a) Fourier Ptychography (FP) using 69 measurements, (b) traditional qDPC and (c) learned designs, for the case of 4, 3, and 2 measurements. Error maps show the difference from the FP reconstruction. (d) Cross-sections show that phase from our learned designs (long-dashed red) is closer to that of FP (solid blue) than traditional qDPC (short-dashed green).}
    \label{fig:fig7}
\end{figure*}

Traditional qDPC uses 4 measurements to adequately cover frequency space. Our learned designs are more efficient and may require fewer measurements; hence, we show learned designs for the cases of 4, 3 and 2 measurements. The designs and their corresponding phase WOTFs are shown in Fig.~\ref{fig:fig4}.  

Comparing our learned designs with previous work, Fig.~\ref{fig:fig5} shows the phase reconstruction for a single simulated test example using 4, 3 and 2 measurements. The ground truth phase is compared with the phase reconstructed using traditional qDPC designs~\cite{Tian:2015fs}, annular illumination designs~\cite{Tian:2015fs}, condition number optimized designs~\cite{marechal2009}, A-optimal designs~\cite{WONG1984295}, and our physics-based learned designs. Table~\ref{table:dBimprovement} reports the peak SNR (PSNR) statistics (mean and standard deviation) for the phase reconstructions from $\mathcal{R}$ evaluated on our set of testing examples. Our learned designs give significant improvement, recovering both the high and low frequencies more accurately. 

\subsection{Experimental Validation}
\label{ssec:expvalid}

To demonstrate that our learned designs generalize well in the experimental setting, we implement our method on an LED array microscope. A commercial Nikon TE300 microscope is equipped with a custom quasi-Dome~\cite{Phillips:17} illumination system (581 programmable RGB LEDs: $\lambda_R = 625$ nm, $\lambda_G = 532$ nm, $\lambda_B = 450$ nm) and a PCO.edge 5.5 monochrome camera ($2560\times2160$, $6.5\mu m$ pixel pitch, 16 bit). We image two samples: a USAF phase target (Benchmark Technologies) and fixed 3T3 mouse fibroblast cells (prepared as detailed in the supplement). In order to validate our method, we compare results against phase experimentally estimated via pupil-corrected Fourier Ptychography (FP)~\cite{Zheng:2013gq,Ou:2014ea,Tian:2014wv} with equivalent resolution. FP is expected to have good accuracy, since it uses significantly more measurements (69 single-LED measurements) and a non-linear reconstruction process.

Using the USAF target, we compare phase reconstructions from FP with traditional qDPC and our learned design measurements (Fig.~\ref{fig:fig6}). Traditional qDPC reconstructions consistently under-estimate the phase values. However, phase reconstructions using our learned design measurements are similar to phase estimated with FP. As the number of measurements is reduced, the performance quality of the reconstruction using traditional qDPC degrades, while the reconstruction using the learned design remains accurate.

To demonstrate our method with live biological samples, we repeated the experiments with 3T3 mouse fibroblast cells. Figure~\ref{fig:fig7} shows that phase reconstructions from traditional qDPC again consistently under-estimate phase values, while phase reconstructions using learned design measurements match the phase estimated with FP well.

\section{Discussion}
\label{sec:discussion}

Our proposed experimental design method efficiently learns the coded-illumination designs by incorporating both the system physics and the non-linear nature of iterative phase recovery. Learned designs with only 2 measurements can efficiently reconstruct phase with quality similar to Fourier Ptychography ($69$ measurements) and better than qDPC ($4$ measurements), giving an improvement in temporal resolution by a factor of 2$\times$ over traditional qDPC and far fewer than FP. Additionally, we demonstrate (Table~\ref{table:dBimprovement}) that the performance of our designs on a set of testing examples is superior to previously-proposed coded-illumination designs. Visually, our learned design reconstructions closely resemble the ground truth phase, with both low-frequency and high-frequency information accurately recovered.

By parameterizing our learning problem with only a few weights per measurement, our method can efficiently learn an experimental design with a small simulated dataset. This enables fast training and reduces computing requirements significantly. Obtaining large experimental datasets for training may be difficult in microscopy, so it is important that our method can be trained on simulated data only. Experimental results in Sec.~\ref{ssec:expvalid} show similar quality to simulated results, with both using the designs learned from simulated data only.

Finally, phase recovery with the learned designs' measurements are trained with a given number of reconstruction iterations (\textit{e.g.} determined by a CPU budget). This makes our method particularly well-suited for real-time processing. qDPC can also be implemented in real-time, but limiting the compute time for the inverse problem (by restricting the number of iterations) limits convergence and causes low-frequency artifacts. Our learned designs incorporate the number of iterations (and hence processing time) into the design process, producing high-quality phase reconstructions within a reasonable compute time.

\section{Outlook}
\label{sec:outlook}

Our method is general to the problem of experimental design. Similar to QPI, many fields (\textit{e.g.} Magnetic resonance imaging (MRI), fluorescence microscopy) use physics-based non-linear iterative reconstruction techniques to achieve state-of-the-art performance. With the correct model parameterization and physically-relevant constraints, our method could be applied to learn optimal design for these applications (\textit{e.g.} undersampling patterns for compressed sensing MRI~\cite{lustig2007sparse}, PSFs for fluorescence microscopy~\cite{Pavani2995}).

Requirements for applying our method are simple: the reconstruction algorithm's updates must be differentiable (\textit{e.g.} gradient update and proximal update) so that analytic gradients of the learning loss can be computed with respect to the design parameters. Of practical importance, the proximal operator of the regularizer should be chosen so that it has a closed form. While this is not a strict requirement, if the operator itself requires an additional iterative optimization, error will have to be backpropagated through an excessive number of iterations. Here, we choose to penalize anisotropic TV, whose proximal operator can be approximated in closed form~\cite{Kamilov:2016gc}. Further, including an acceleration update improves the convergence of gradient-based reconstructions. As a result, the \textit{unrolled network} can be constructed using fewer layers than its unaccelerated counterpart. This will reduce both computation time and training requirements.

\section{Conclusion}
\label{sec:conc}

We have presented a general framework for incorporating the non-linearities of regularized reconstruction and known system physics to learn optimal experimental design. Here, we have applied this method to learn coded-illumination source designs for quantitative phase recovery. Our coded-illumination designs can improve the temporal resolution of the acquisition and enable real-time processing, while maintaining high accuracy. We demonstrated here that our learned designs achieve high-quality reconstructions experimentally without the need for retraining.

\section*{Funding Information}

This work was supported by STROBE: A National Science Foundation Science \& Technology Center under Grant No. DMR 1548924 and by the Gordon and Betty Moore Foundation's Data-Driven Discovery Initiative through Grant GBMF4562 to Laura Waller (UC Berkeley). Laura Waller is a Chan Zuckerberg Biohub investigator. Michael R. Kellman is additionally supported by the National Science Foundation's Graduate Research Fellowship under Grant No. DGE 1106400. Emrah Bostan’s research is supported by the Swiss National Science Foundation (SNSF) under grant P2ELP2 172278.

\section*{Acknowledgment}

The authors would like to thank Professor Michael Lustig for his guidance and advice.

\ifCLASSOPTIONcaptionsoff
  \newpage
\fi



\bibliographystyle{IEEEtran}
\bibliography{LDPC2}

\vfill


\end{document}